# Full-resolution MLPs Empower Medical Dense Prediction

Mingyuan Meng[1,2], Yuxin Xue[1], Dagan Feng[1], Lei Bi[2], and Jinman Kim[1]

[1] School of Computer Science, The University of Sydney, Australia.
[2] Institute of Translational Medicine, Shanghai Jiao Tong University, China.

## Abstract

*Dense prediction is a fundamental requirement for many medical vision tasks such as medical image restoration, registration, and segmentation. The most popular vision model, Convolutional Neural Networks (CNNs), has reached bottlenecks due to the intrinsic locality of convolution operations. Recently, transformers have been widely adopted for dense prediction for their capability to capture long-range visual dependence. However, due to the high computational complexity and large memory consumption of self-attention operations, transformers are usually used at downsampled feature resolutions. Such usage cannot effectively leverage the tissue-level textural information available only at the full image resolution. This textural information is crucial for medical dense prediction as it can differentiate the subtle human anatomy in medical images. In this study, we hypothesize that Multi-layer Perceptrons (MLPs) are superior alternatives to transformers in medical dense prediction where tissue-level details dominate the performance, as MLPs enable long-range dependence at the full image resolution. To validate our hypothesis, we develop a full-resolution hierarchical MLP framework that uses MLPs beginning from the full image resolution. We evaluate this framework with various MLP blocks on a wide range of medical dense prediction tasks including restoration, registration, and segmentation. Extensive experiments on six public well-benchmarked datasets show that, by simply using MLPs at full resolution, our framework outperforms its CNN and transformer counterparts and achieves state-of-the-art performance on various medical dense prediction tasks.*

## 1. Introduction

Dense prediction, also known as pixel-wise prediction, is a fundamental problem in computer vision. It aims to learn pixel-wise mapping from input images to various complex output structures, including segmentation, depth estimation, and image restoration [1]. For medical images, dense prediction is also a fundamental requirement for a wide range of medical vision tasks such as deformable medical image registration, medical image restoration, and organ/lesion segmentation. Compared to classification and recognition problems, dense prediction performs pixel-wise classification or regression on the whole image. This raises the requirement for high-resolution contextual information in hierarchical multi-scale image features, thus incurring additional challenges for model design [2].

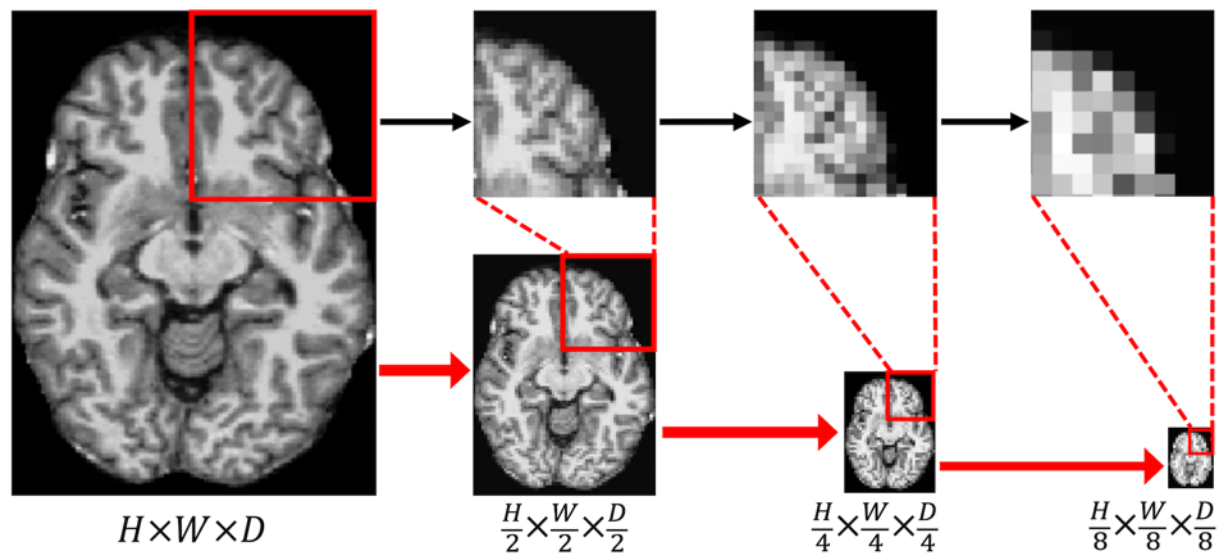

Figure 1: Illustration of medical images after downsampling. A 2D slice of a 3D brain MRI image is presented, where the brain anatomy cannot be well identified in the downsampled images.

Convolutional Neural Networks (CNNs), as the most popular model in computer vision for the past decade, have been widely adopted for medical dense prediction, where U-Net [3] or its variants [4-6] are commonly used as the backbone with a hierarchical pyramid encoder-decoder structure. However, despite their popularity, CNNs have met bottlenecks due to the intrinsic locality of convolution operations. CNNs tend to increase the receptive field by progressively downsampling image features and stacking convolutional layers [7]. Unfortunately, in this manner, the receptive fields in high resolutions are limited and thus cannot capture fine-grained long-range visual dependence. The long-range dependence has been proven to be crucial for dense prediction as it provides important contextual information for semantics understanding [1].

Recently, transformers and their hierarchical variants (e.g., Swin transformer [8]) have been widely adopted as alternatives to CNNs in medical dense prediction for their great capability to capture long-range dependence via self-attention [9-12]. By employing local window-based [8, 13] or dimension-reduced/pooled [14, 15] self-attention in the hierarchical pyramid structure, transformers can process finer-grained features to capture long-range dependence at multiple scales, enabling them to achieve better dense prediction performance than CNNs [10-12]. Nevertheless, self-attention incurs high computational complexity and large memory consumption, especially for 3D medical images, such as Magnetic Resonance Imaging (MRI), Positron Emission Tomography (PET), and Computed Tomography (CT). Therefore, transformers are usually employed at downsampled feature resolutions [8-16] and thus cannot effectively leverage the tissue-level textural information available only at the full image resolution.

The tissue textural information is especially crucial for medical dense prediction. Compared with natural images, medical images usually delineate a certain body region of different patients. Therefore, they are highly similar to each other and contain many subtle anatomical and pathological characteristics that cannot be differentiated without tissue-level textural information. As exemplified in Figure 1, brain MRI images contain complex anatomical structures such as the cerebral cortex, and these anatomical details have been lost after downsampling. This information loss is acceptable for image-wise prediction tasks as the textural information can be encoded into high-level (downsampled) features to characterize image semantics. However, this is catastrophic for medical dense prediction, especially for challenging tasks, e.g., medical image restoration aiming to obtain images with sufficient anatomical details or deformable image registration aiming to find precise pixel-wise spatial correspondence between anatomical structures. To compensate for this information loss, convolutional layers were used at the full/half image resolution in hybrid CNN-transformer models for medical dense prediction [10, 12]. Nevertheless, the fine-grained long-range dependence at the full/half resolution still cannot be captured.

The reliance on modeling long-range dependence has motivated another research trend toward models based on Multi-layer Perceptrons (MLPs) [17]. By removing self-attention, MLPs further reduce the inductive bias and, more importantly, they are more computationally efficient than transformers while also being able to capture long-range dependence [18-22]. Recently, MLPs have been used in the hierarchical pyramid structure and achieved comparable performance with state-of-the-art methods based on CNNs and/or transformers [23-28]. Due to the high efficiency in computation and memory consumption, MLPs have the potential to be used at full image resolution to capture fine-grained long-range dependence [23]. However, current studies still tend to employ MLPs beginning from the 1/4 image resolution after 4×4 patch embedding [24-28]. This is likely because existing MLP-based models were mainly designed and evaluated for natural vision tasks where detailed textural information is not indispensable and high-level semantic information dominates the performance.

In this study, we investigate the unnoticed potential of MLPs on capturing full-resolution longe-range dependence for medical dense prediction. To this end, we develop a full-resolution hierarchical MLP framework that employs MLP blocks in a hierarchical manner to extract multi-scale features beginning from the full image resolution. The extracted features are leveraged for a variety of medical dense prediction tasks by using different task-specific decoders, including medical image restoration, deformable medical image registration, and organ/lesion segmentation. Moreover, we evaluate various MLP blocks (e.g., Hire-MLP [24], Sparse MLP [26], etc.) in our MLP framework. Experiments on six widely benchmarked public datasets show that, by simply employing MLPs at the full image resolution, our MLP framework consistently outperforms its CNN and transformer counterparts regardless of which MLP block is adopted. We also identify that our framework achieves better performance than the existing methods optimized for one single medical dense prediction task.

We hope that our study can inspire new discussions on the employment of MLPs in medical vision tasks and motivate the research community to rethink the potential of MLPs as a valuable alternative to CNNs and transformers for medical dense prediction.

## 2. Related Work

### 2.1. Multi-layer Perceptrons (MLPs)

MLP-based models have attracted much attention in the vision community due to their capability to capture long-range visual dependence without relying on self-attention. Inherited from the earlier Visual Transformer (ViT) [29], early MLP-based models typically process single-scale image features following 16×16 patch embedding [18-22]. For instance, Tolstikhin et al. [18] proposed an MLP-Mixer that employs MLPs to separately mix channel and spatial information via matrix transpose. Liu et al. [19] then proposed an MLP-Mixer variant (gMLP) by introducing spatial gating units for spatial projections. These models achieved promising performance in image classification. However, they still have a large gap from being a general vision backbone, especially for dense prediction tasks that tend to leverage hierarchical multi-scale image features.

Recently, MLP-based models have been designed as hierarchical pyramid structures [24-28], which extends the applicability of MLPs to dense prediction. For example, Guo et al. [24] proposed a hierarchical MLP-based model (Hire-MLP) via hierarchical rearrangement. Chen et al. [25] proposed a hierarchical MLP-based model, CycleMLP, by introducing a cycle fully connected layer to mix spatial information. Tang et al. [26] proposed a hierarchical sparse MLP model (sMLP) that mixes spatial information along each axial direction (horizontal and vertical). These models employed MLPs beginning from the 1/4 image resolution after 4×4 patch embedding, which hinders them from capturing the fine-grained long-range dependence at the full and half image resolution.

We also notice that Tu et al. [23] proposed a hierarchical MLP-based model (MAXIM) for image processing, where MLP blocks are used to process full-resolution features. This model attained state-of-the-art performance on natural image processing tasks such as denoising and deblurring, which hints at the importance of fine-grained long-range dependence for dense prediction. However, Hu et al. [23]'s study is limited to natural image processing tasks, and the MAXIM is difficult to generalize to 3D medical dense prediction due to its complex architecture requiring large GPU memory. More importantly, Hu et al. [23] focused on designing a specialized image processing method with a novel multi-axis gated MLP block, so that they did not identify and validate the improvements brought from the full-resolution MLPs. So far, the employment of full-resolution MLPs has not been extensively investigated, especially for medical dense prediction.

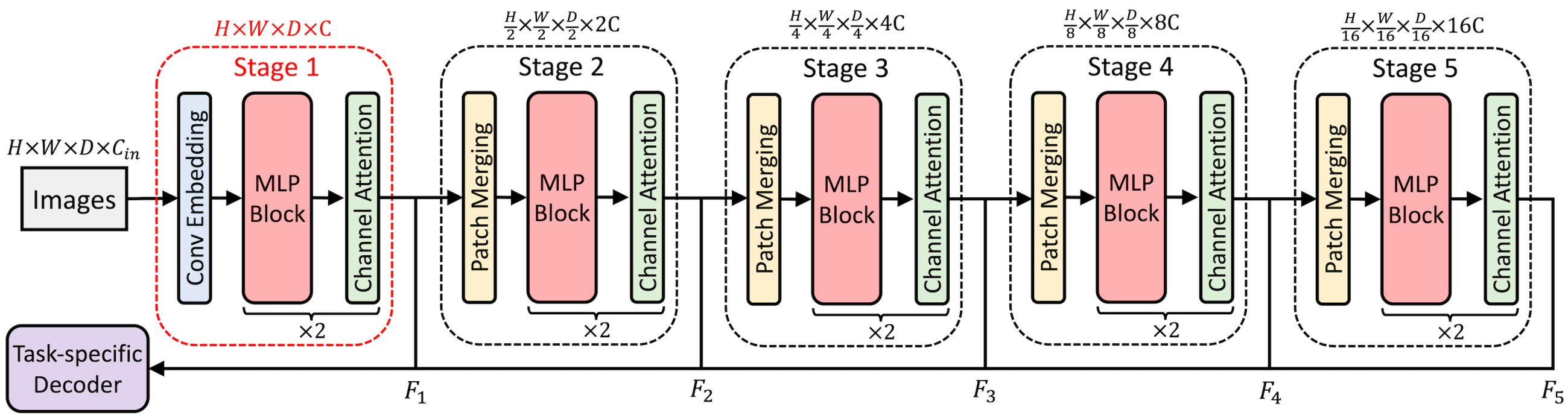

Figure 2: The architecture of full-resolution hierarchical MLP framework. This framework extracts multi-scale pyramid image features beginning from the full image resolution (Stage 1), which serves as a general visual backbone for various medical dense prediction tasks.

### 2.2. Medical Dense Prediction

Medical dense prediction includes a wide range of pixel-wise prediction tasks, e.g., medical image restoration [30], registration [31], segmentation [32], and processing [33]. Since the introduction of U-net [3, 4], its variants have been widely used in medical dense prediction tasks as they can leverage multi-scale features via a hierarchical encoder-decoder design and recover fine-grained detail information via skip connections [5, 6, 34-36]. Recently, transformers, especially Swin transformers [8], have been introduced for medical dense prediction and attained state-of-the-art performance [9-12, 37-39]. To leverage the high-resolution tissue textural information that cannot be processed by transformers, hybrid CNN-transformer models were used [9, 10, 12, 37], where convolutional layers are employed at high resolutions (full/half image scale) while transformers are employed to process the downsampled features after patch embedding. These models cannot capture the fine-grained long-range dependence at high resolutions.

The employment of MLPs for medical dense prediction has not been extensively investigated, and most existing studies focused on medical image segmentation [40-43]. For example, Valanarasu et al. [40] proposed a rapid MLP-based medical image segmentation network (UNeXt) that uses tokenized shifted MLPs at the bottleneck of U-net. Shi et al. [41] proposed to use CycleMLP as the encoder for polyp segmentation. In addition to segmentation, there are only a few preliminary investigations on other medical dense prediction tasks [44, 45]. Wang et al. [44] proposed a registration model that employs MLP-Mixer blocks for 2D echocardiography registration. Mansour et al. [45] also proposed a denoising model using MLP-Mixer blocks and evaluated it with 2D knee MRI slices.

These MLP-based medical dense prediction models tend to follow the network structure adopted in the earlier MLP studies [18, 24-27], where MLP blocks are employed after patch embedding. Therefore, they cannot effectively leverage the crucial tissue-level textural information and have limited capabilities in capturing fine-grained long-range dependence at full resolution. Moreover, these MLP-based models were particularly optimized for individual tasks and have limited generalizability to other medical dense prediction tasks. It also should be noted that, unlike previous studies that optimize MLP-based models for individual tasks, our study shows that simply using MLPs at full resolution has already produced consistently great performance in various medical dense prediction tasks.

## 3. Experimental Method

### 3.1. Full-resolution Hierarchical MLPs

To validate the effectiveness of employing MLPs at full resolution, we develop a full-resolution hierarchical MLP framework as shown in Figure 2. This framework has five MLP stages to extract multi-scale image features, [$F_1$, $F_2$, $F_3$, $F_4$, $F_5$], from the full to the 1/16 image resolution. Each MLP stage contains two MLP blocks and channel attention blocks to process features at each resolution. For image embedding, convolutional embedding is adopted to obtain initial image features at the full resolution, and then patch merging layers are adopted between stages to downsample the feature resolution and increase the feature dimension.

This framework can serve as a general visual backbone (encoder) for medical dense prediction tasks, where the extracted image features $[F_1, F_2, F_3, F_4, F_5]$ are leveraged for various tasks using task-specific decoders.

**Convolutional Embedding.** We discard the common patch embedding approach [29] and adopt a non-strided 3×3×3 convolutional layer for image embedding, which converts the input images $I \in \mathbb{R}^{H\times W\times D\times C_{in}}$ into the initial image features $F_{init} \in \mathbb{R}^{H\times W\times D\times C}$. The $C$ is the embedding dimension at the first stage, which is set as 24.

**MLP Block.** Our framework is applicable to a variety of existing MLP blocks such as Hire-MLP [24], sparse MLP [26], and multi-axis gated MLP [23]. We adopt the multi-axis gated MLP [23] as default and also follow [23] to employ a residual channel attention block after each MLP block to highlight important feature channels. This MLP block is chosen for its capability to capture both local and global information by regional and dilated MLP operations. Nevertheless, there are experiments (section 5.4) showing that different MLP blocks performed similarly in our framework. The detailed settings of MLP blocks are provided in the supplementary materials.

### 3.2. Task-specific Decoders

The full-resolution hierarchical MLP framework is used for three medical dense prediction tasks, including medical image restoration, registration, and segmentation, via task-specific decoders. We adopt well-established decoders for a fair comparison with existing medical dense prediction methods. For segmentation and restoration, we adopt the decoder of Swin-UNETR [12] (denoted by MLP-Unet). For registration, we adopt the decoder of TransMorph [10] (denoted by MLPMorph). The network architecture and detailed architectural settings of task-specific decoders are presented in the supplementary materials.

## 4. Experimental Setup

### 4.1. Evaluation Tasks and Datasets

Our MLP framework was evaluated on the tasks of medical image restoration, registration, and segmentation, with six well-established public benchmark datasets.

**Restoration.** We restored low-dose PET images into standard-dose PET images. Standard-dose PET images are acquired with a standard dose of radiation tracer, while low-dose PET images are acquired with lower radiation dosages [47]. This task has attracted wide attention [48-53] as it enables lower radiation dosage in PET imaging and thus reduces the risk of radiation exposure. We adopted two public datasets from the Ultra-Low Dose Imaging (Ultra 2022) challenge [54], including 209 sets of images acquired from Siemens Biograph Vision Quadra (SBVQ) scanners and 320 sets of images scanned by United Imaging uEXPLORER (UIU) scanners. Each dataset was randomly divided for training, validation, and testing with a ratio of 70%, 20%, and 10%. Each set of images contains six whole-body PET scans of a patient, including five low-dose images with different radiation dosages (i.e., with different noise levels) and a standard-dose image as the ground truth label. The image with the lowest radiation dosage (i.e., with the most noise) is called the ultra-low dose PET image, serving as the most challenging case.

**Registration.** We registered brain MRI images acquired from different patients, which is a well-benchmarked task for deformable medical image registration [6, 10, 55-59]. 2,656 brain MRI images acquired from four unlabeled public datasets (ADNI [60], ABIDE [61], ADHD [62], and IXI [63]) were adopted for training. Two public brain MRI datasets with anatomical segmentation labels (Mindboggle [64] and Buckner [65]) were used for validation and testing. The Mindboggle dataset contains 100 MRI images and was randomly split into 50/50 images for validation/testing. The Buckner dataset contains 40 MRI images and was used for independent testing only.

**Segmentation.** We segmented head and neck tumors in PET-CT images and cardiac structures in MRI images, both of which are widely benchmarked tasks for medical image segmentation [9, 11, 66-69]. For head and neck tumor segmentation, we adopted a public dataset from the HEad and neCK TumOR segmentation (HECKTOR 2022) challenge [66], including 488 head and neck PET-CT images with ground truth labels of primary tumors and metastatic lymph nodes. For cardiac image segmentation, we adopted the public ACDC dataset [67], including 150 cardiac MRI images with ground truth labels of left ventricles, right ventricles, and myocardium. These two datasets were randomly split for training, validation, and testing with a ratio of 70%, 20%, and 10%.

Additional details about the datasets and preprocessing procedures are provided in the supplementary materials.

### 4.2. Comparison Methods

Our MLP framework was extensively compared with existing methods, including the state-of-the-art methods specialized for the three evaluation tasks.

For image restoration, the comparison methods consist of 3D-UNet [48], 3D-cGAN [49], Stack-GAN (stacked 3D-cGANs) [49], Cycle-GAN [50], AR-GAN [51], Swin-UNETR [12], and SS-AEGAN [52]. The SS-AEGAN is one of the top-performing methods at the Ultra 2022 challenge. The Swin-UNETR was adapted to restoration by modifying its output layer.

For image registration, the comparison methods include VoxelMorph [6], TransMorph [10], LKU-Net [56], Swin-VoxelMorph [57], TransMatch [58], and MLP-Mixer [44].

For image segmentation, the comparison methods are 3D-Unet [4], TransUnet [9], Swin-Unet [11], UNETR [37], Swin-UNETR [12], and UNeXt [40].

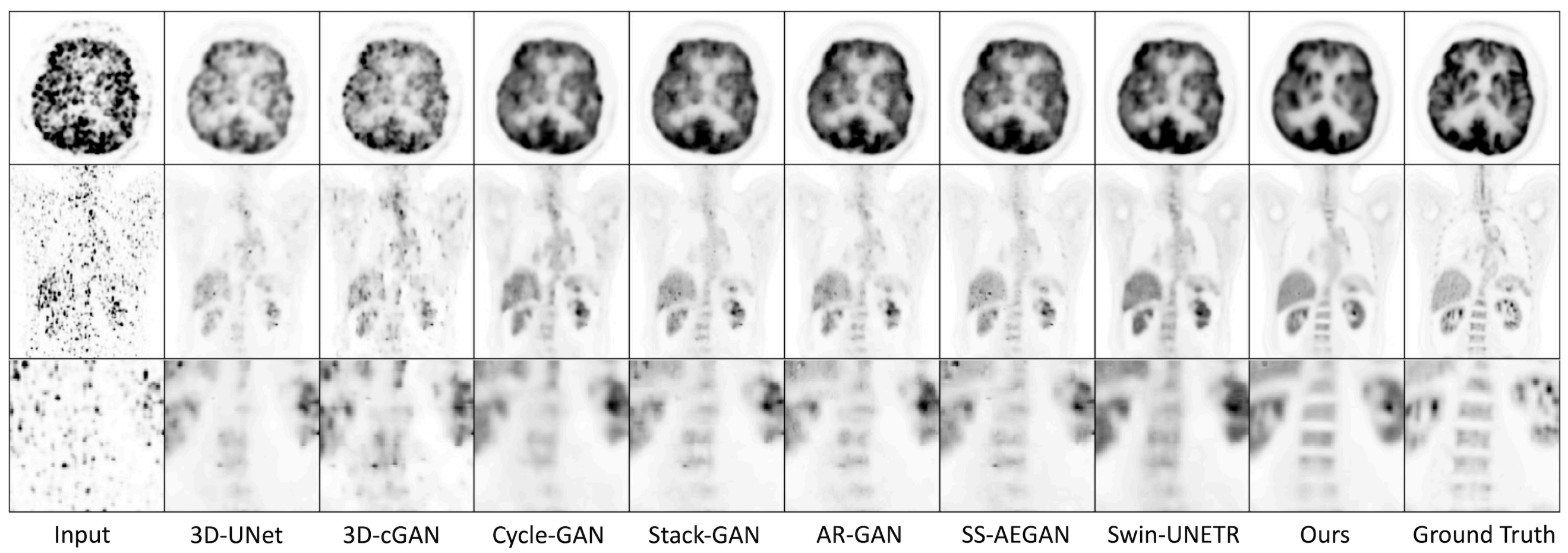

Figure 3: Qualitative comparison for image restoration. The input is an ultra-low dose PET image, while the ground truth is the standard-dose PET image. The brain, body, and spine regions are zoomed in and presented in the first, second, and third rows, respectively.

| Method | UIU dataset | | | | | | SBVQ dataset | | | | | |
|---|---|---|---|---|---|---|---|---|---|---|---|---|
| | PSNR (dB) ↑ | | SSIM ↑ | | NRMSE (%) ↓ | | PSNR (dB) ↑ | | SSIM ↑ | | NRMSE (%) ↓ | |
| | Ultra-low | Avg | Ultra-low | Avg | Ultra-low | Avg | Ultra-low | Avg | Ultra-low | Avg | Ultra-low | Avg |
| Before restoration | 43.014 | 48.464 | 0.9702 | 0.9892 | 1.042 | 0.6288 | 43.046 | 48.023 | 0.9965 | 0.9976 | 0.960 | 0.606 |
| 3D-UNet [48] | 47.645 | 50.323 | 0.9810 | 0.9900 | 0.577 | 0.402 | 48.519 | 49.932 | 0.9915 | 0.9939 | 0.475 | 0.404 |
| 3D-cGAN [49] | 49.086 | 50.731 | 0.9907 | 0.9946 | 0.462 | 0.386 | 49.710 | 51.200 | 0.9916 | 0.9939 | 0.414 | 0.357 |
| Cycle-GAN [50] | 49.347 | 51.533 | 0.9924 | 0.9955 | 0.444 | 0.372 | 49.870 | 51.872 | 0.9910 | 0.9941 | 0.420 | 0.344 |
| Stack-GAN [49] | 50.039 | 51.504 | 0.9941 | 0.9957 | 0.419 | 0.341 | 50.839 | 52.228 | 0.9931 | 0.9939 | 0.378 | 0.309 |
| AR-GAN [51] | 50.327 | 53.340 | 0.9930 | 0.9954 | 0.413 | 0.298 | 51.265 | 53.706 | 0.992 | 0.9932 | 0.346 | 0.266 |
| SS-AEGAN [52] | 50.824 | 53.854 | 0.9941 | 0.9965 | 0.403 | 0.288 | 51.468 | 53.858 | 0.9937 | 0.9957 | 0.338 | 0.263 |
| Swin-UNETR [12] | 51.477 | 53.828 | 0.9913 | 0.9969 | 0.339 | 0.264 | 51.875 | 53.623 | 0.9995 | 0.9996 | 0.315 | 0.264 |
| MLP-Unet (Ours) | **52.695** | **54.838** | **0.9999** | **0.9999** | **0.297** | **0.235** | **52.508** | **54.433** | **0.9997** | **0.9997** | **0.296** | **0.242** |

Table 1: Quantitative comparison for image restoration. The best result in each column is in bold. ↑: higher is better. ↓: lower is better. Ultra-low: the results of ultra-low dose PET images. Avg: average results of five low-dose PET images with different radiation dosages.

## 4.3. Evaluation Metrics

We strictly followed the standard evaluation metrics used in the widely recognized challenge competitions [54, 66, 70, 71] to ensure comprehensive evaluation and a fair comparison with existing methods.

For image restoration, normalized root mean squared error (NRMSE), peak signal-to-noise ratio (PSNR), and structural similarity index measurement (SSIM) were used to evaluate the similarity between the restored images and ground truth labels (standard-dose PET images). All five dosage levels of low-dose PET images were restored and we calculated their weighted average results following the evaluation scheme at the Ultra 2022 challenge [54].

For image registration, the registration accuracy was evaluated using the Dice similarity coefficients (DSC) between the segmentation labels of the registered images. The smoothness and invertibility of the produced spatial transformations were evaluated via the percentage of negative Jacobian determinants (NJD) [72]. The inference time was used to evaluate registration speed. Registration methods are required to perform fast, accurate registration while also keeping the transformation smoothness.

For image segmentation, DSC and Hausdorff Distance at the 95th percentile (HD95) were adopted to evaluate the consistency between the segmentation predictions and manually annotated labels. The mean results calculated on all labeled regions/structures are reported.

The detailed definitions of these evaluation metrics are provided in the supplementary materials.

## 4.4. Experimental Designs

Our experiments were conducted using PyTorch on an NVIDIA A100 GPU with 40GB memory. The detailed training procedures are described in the supplementary materials, including the training schemes (e.g., learning rate, optimizer) and loss functions used for three evaluation tasks. Our implementation code is is publicly available at *https://github.com/MungoMeng/DensePred-FullMLP*.

For a fair comparison, all the comparison methods were implemented following the same training schemes and loss functions as ours, and their feature dimensions were chosen under the constraints of GPU memory. Moreover, the 2D comparison methods were reimplemented as 3D models by replacing all 2D operations with 3D counterparts.

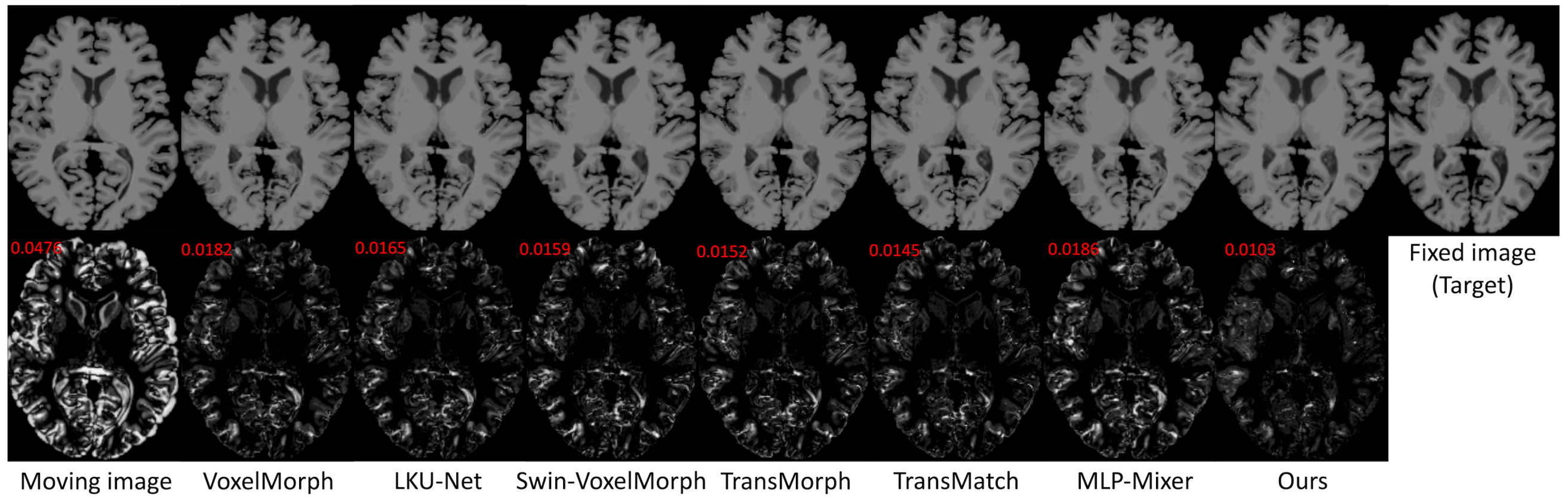

Figure 4: Qualitative comparison for registration. Below each image is an error map that shows the intensity differences from the fixed image (target), with the mean absolute error placed in the upper left corner. A cleaner error map indicates a better registration result.

| Method | | Mindboggle dataset | | Buckner dataset | | Inference time | |
|---|---|---|---|---|---|---|---|
| | | DSC ↑ | NJD (%) ↓ | DSC ↑ | NJD (%) ↓ | CPU (s) | GPU (s) |
| Before registration | | 0.347 | / | 0.406 | / | / | / |
| VoxelMorph [6] | CNN-based | 0.565 | 2.383 | 0.595 | 2.229 | **3.88** | **0.16** |
| LKU-Net [56] | CNN-based | 0.578 | 2.123 | 0.614 | 2.036 | 4.34 | 0.19 |
| Swin-VoxelMorph [57] | Transformer-based | 0.576 | 2.026 | 0.614 | 1.869 | 5.96 | 0.34 |
| TransMorph [10] | Hybrid CNN-Transformer | 0.582 | 2.243 | 0.615 | 2.085 | 4.84 | 0.23 |
| TransMatch [58] | Hybrid CNN-Transformer | 0.585 | 1.920 | 0.620 | 1.804 | 4.19 | 0.18 |
| MLP-Mixer [44] | MLP-based | 0.557 | 2.139 | 0.589 | 2.026 | 4.43 | 0.20 |
| MLPMorph (Ours) | Full-resolution MLP | **0.606** | **1.863** | **0.635** | **1.772** | 5.12 | 0.26 |

Table 2: Quantitative comparison for image registration. The best result in each column is in bold. ↑: higher is better. ↓: lower is better.

In addition to the comparison with existing methods, we also performed two ablation studies to further validate the effectiveness of full-resolution MLPs. In the first ablation study, we gradually replaced the MLP block at each stage with a Swin transformer block [8] or residual convolution block [46], and then observed the performance degradation. In the second ablation study, we tried various MLP blocks in our MLP framework to evaluate the impacts of choosing different MLP blocks on the prediction performance. The detailed experimental settings of these ablation studies are provided in the supplementary materials.

## 5. Results and Discussion

### 5.1. Restoration Performance

Table 1 presents a quantitative comparison among our MLP-Unet and existing methods for low-dose PET image restoration. The results of ultra-low dose PET images are reported separately in addition to the average results of five dosage levels, which shows the experimental results at the most challenging conditions. The Swin-UNETR, as a Swin transformer-based method, achieved better performance than all other CNN-based comparison methods, confirming the benefits of capturing long-range dependence. However, the Swin-UNETR employs swin transformers beginning from the half image resolution and, thereby, cannot capture the long-range dependence at full resolution. By using MLPs at full resolution, our MLP-Unet overcame this limitation and achieved the best performance on all three evaluation metrics on both datasets. Moreover, we found that, compared to the average results, both Swin-UNETR and our MLP-Unet achieved larger improvements over other methods in ultra-low dose PET images. This is likely because ultra-low dose PET images are extremely sparse in intensity (exemplified in Figure 3), where the small receptive fields of convolutional layers are incapable of capturing sufficient information that allows for accurate restoration. The Swin-UNETR enlarges the receptive field via Swin transformers, but it still relies on convolutional layers to process full-resolution features. Our MLP-Unet, in contrast, has a large receptive field even at full resolution, which enables it to capture more textural detail information and gain larger improvements than the Swin-UNETR.

Figure 3 provides a qualitative comparison among our MLP-Unet and existing methods. Consistent with the quantitative results (Table 1), the restoration result of our MLP-Unet is visually the most similar to the ground truth standard-dose PET image. It should be noted that our MLP-Unet has also improved the restoration of the anatomical structure on the PET images, such as the brain and spine, while comparison methods resulted in poorer restoration.

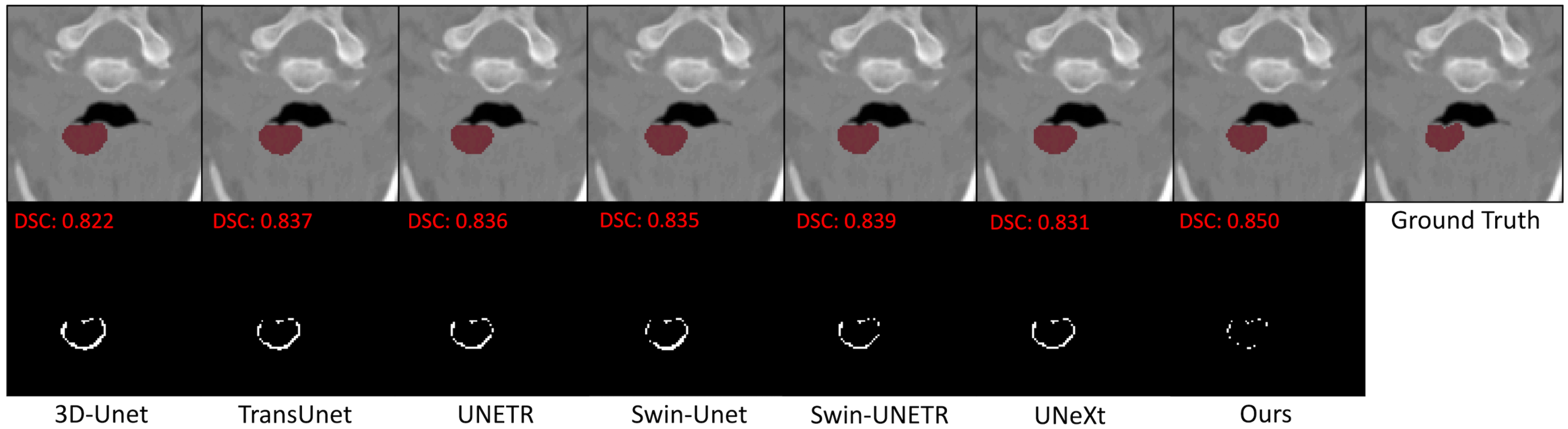

Figure 5: Qualitative comparison for image segmentation. Below each image is an error map that shows the differences between the segmented regions and the ground truth label. The segmented tumor regions are highlighted in red.

| Method | HECKTOR dataset | | ACDC dataset | |
|---|---|---|---|---|
| | DSC ↑ | HD95 ↓ | DSC ↑ | HD95 ↓ |
| 3D-Unet [4] | 0.743 | 19.856 | 0.799 | 6.667 |
| TransUnet [9] | 0.750 | 18.402 | 0.802 | 6.324 |
| UNETR [37] | 0.763 | 14.742 | 0.815 | 5.246 |
| Swin-Unet [11] | 0.760 | 18.588 | 0.829 | 4.214 |
| Swin-UNETR [12] | 0.765 | 17.177 | 0.832 | 4.502 |
| UNeXt [40] | 0.750 | 17.943 | 0.801 | 5.589 |
| MLP-Unet (Ours) | **0.778** | **12.959** | **0.842** | **3.557** |

Table 3: Quantitative comparison for segmentation. The best results are in bold. ↑: the higher is better. ↓: the lower is better.

## 5.2. Registration Performance

Table 2 and Figure 4 present quantitative and qualitative comparisons among our MLPMorph and existing methods for deformable brain MRI registration. The quantitative and qualitative results both show that our MLPMorph outperformed all the comparison methods. In Figure 4, the registration result of our MLPMorph is the most consistent with the fixed image, thus resulting in the cleanest error map. In Table 2, our MLPMorph also achieved the highest registration accuracy (DSC) while not sacrificing the smoothness of the predicted spatial transformations (NJD). In addition, the inference time in Table 2 shows that our MLPMorph has a similar registration speed to the existing deep registration methods.

Both the Swin-VoxelMorph and TransMorph achieved higher DSCs than their CNN counterpart (VoxelMorph), showing the benefits of modeling long-range dependence via transformers. It should be noted that the TransMorph achieved higher DSCs than the Swin-VoxelMorph. The Swin-VoxelMorph uses Swin transformers after 4×4×4 patch embedding and overlooks the image features at the full/half image resolution, while the TransMorph adopts additional convolutional layers to process full/half-scale image features. This demonstrates that detailed textural information at high resolution is crucial for deformable medical image registration. Nevertheless, the TransMorph cannot effectively leverage this full-resolution information due to the intrinsic locality of convolution operations, and we attribute this to its lower performance than MLPMorph. In addition, the method based on MLP-Mixer [44] had poor performance. We attribute this to the fact that this method adopts MLP-Mixer to perform registration at the 1/4, 1/8, and 1/16 image resolutions and thus fails to capture fine-grained long-range dependence at high resolution.

## 5.3. Segmentation Performance

Table 3 presents a quantitative comparison among our MLP-Unet and existing methods for medical segmentation. Consistent with the restoration and registration results, the transformer-based methods outperformed the CNN-based 3D-Unet, and the hybrid Swin-UNETR outperformed the pure transformer-based Swin-Unet by using convolutional layers to process full-scale features. These results further validate that the modeling of long-range dependence and high-resolution textural information is crucial for medical dense prediction. By modeling the fine-grained long-range dependence among full-resolution textural information, our MLP-Unet attained the best segmentation performance over all the comparison methods. The MLP-based UNeXt failed to outperform the transformer-based methods as it uses MLP blocks only at the bottleneck of Unet and, hence, does not fully exploit the advantages of MLPs.

We noticed that the improvements of using our MLP framework in segmentation tasks are not as large as in restoration and registration tasks. This is likely because segmentation tasks are reliant more on the high-level semantic information encoded in deep layers to identify segmentation targets. Nevertheless, the high-resolution textural information is still beneficial for segmentation to identify the boundary of target regions. As exemplified in Figure 5, although all methods successfully located the target tumor, our MLP-Unet delineated the closest tumor boundaries to the ground truth label.

| Method | Restoration (UIU, ultra-low dose PET) | | | | | Registration (Mindboggle) | | | | Segmentation (HECKTOR) | | | |
|---|---|---|---|---|---|---|---|---|---|---|---|---|---|
| | PSNR (dB) ↑ | SSIM ↑ | NRMSE (%) ↓ | Mem | Param | DSC ↑ | NJD (%) ↓ | Mem | Param | DSC ↑ | HD95 ↓ | Mem | Param |
| PPPPP | **52.695** | **0.9999** | **0.297** | 32.2GB* | 44.1M | **0.606** | **1.863** | 33.2GB* | 33.9M | **0.778** | **12.959** | 35.6GB* | 44.1M |
| CPPPP | 51.838 | 0.9998 | 0.322 | 24.1GB* | 43.0M | 0.595 | 1.925 | 24.8GB* | 32.8M | 0.772 | 14.458 | 24.8GB* | 43.0M |
| CCPPP | 51.316 | 0.9997 | 0.346 | 23.8GB | 41.8M | 0.586 | 1.993 | 23.4GB | 31.6M | 0.768 | 15.151 | 23.8GB | 41.8M |
| CTTTT | 51.207 | 0.9910 | 0.348 | 34.8GB* | 25.9M | 0.590 | 2.016 | 29.4GB* | 15.7M | 0.765 | 16.280 | 34.2GB* | 25.9M |
| CCTTT | 50.851 | 0.9910 | 0.367 | 24.5GB | 25.8M | 0.584 | 2.154 | 24.6GB | 15.6M | 0.762 | 17.285 | 24.8GB | 25.8M |

Table 4: Ablation study where MLP (P), Swin transformer (T), and convolution (C) blocks are used at five stages of our framework. The PPPPP denotes the full-resolution MLP framework that employs MLP blocks at all five stages. The best result in each column is in bold. Param: Number of learnable parameters. Mem: GPU memory consumption during training. ↑: the higher is better. ↓: the lower is better. *: The checkpointing technique was used to trade computation for memory.

| MLP blocks | Restoration (UIU, ultra-low dose PET) | | | Registration (Mindboggle) | | Segmentation (HECKTOR) | |
|---|---|---|---|---|---|---|---|
| | PSNR (dB) ↑ | SSIM ↑ | NRMSE (%) ↓ | DSC ↑ | NJD (%) ↓ | DSC ↑ | HD95 ↓ |
| Hire-MLP [24] | **52.726** | 0.9998 | **0.295** | 0.595 | **1.773** | **0.780** | 14.654 |
| sMLP [26] | 52.527 | **0.9999** | 0.301 | 0.598 | 1.850 | 0.777 | **12.612** |
| Swin-MLP [8] | 52.515 | **0.9999** | 0.303 | 0.605 | 1.812 | 0.772 | 13.854 |
| Multi-axis gated MLP [23] | 52.695 | **0.9999** | 0.297 | **0.606** | 1.863 | 0.778 | 12.959 |
| Swin-UNETR/TransMorph [10,12] | 51.477 | 0.9913 | 0.339 | 0.582 | 2.243 | 0.765 | 17.177 |

Table 5: Ablation study where various MLP blocks are adopted in our framework. The best result in each column is in bold. ↑: the higher is better. ↓: the lower is better. The results of Swin-UNETR (for restoration/segmentation) and TransMorph (for registration) are presented in the last row as comparison benchmarks.

### 5.4. Ablation Studies

Table 4 presents the ablation study where MLP blocks in our framework were replaced by Swin transformers or convolution blocks. The GPU memory consumed during training and the parameter numbers are also reported in Table 4. Replacing MLP blocks at the first two stages with convolution blocks degraded the performance on all three evaluation tasks, demonstrating the effectiveness of employing MLP blocks at full and half resolution. We attempted to employ the Swin transformer block at the first stage. However, this is infeasible with our current GPU memory (40GB). Therefore, we used Swin transformer blocks beginning from the second or third stage and this produced two hybrid CNN-transformer models. Compared to our full-resolution MLP model, the model with half-resolution transformers consumed similar GPU memory but resulted in lower performance on all three evaluation tasks. Our results demonstrate that the MLPs can be used at higher resolution to capture finer-grained long-range dependence under the same GPU memory constraints. In addition, the two hybrid CNN-transformer models were also outperformed by their hybrid CNN-MLP counterparts, which suggests that MLPs have advantages over Swin transformers even when being used at the same resolution. Compared with Swin transformers, recent hierarchical MLPs, such as the multi-axis gated MLP [23] used in the experiments, allow for larger receptive fields as they mix spatial information via efficient MLP operations instead of relying on heavy self-attention operations.

Table 5 presents the ablation study where various MLP blocks were used in our full-resolution MLP framework. We evaluated four different existing MLP blocks on three evaluation tasks, where the Swin-UNETR (restoration and segmentation) and TransMorph (registration) are regarded as comparison benchmarks. We found that, no matter which MLP block was adopted, the methods using our full-resolution MLP framework consistently outperformed the comparison benchmarks. This finding suggests that the improvements found in this study are not derived from the design of MLP blocks but rather from the novelty of applying them at full resolution.

## 6. Conclusion

We have validated the effectiveness of leveraging MLPs at full resolution for medical dense prediction via a full-resolution hierarchical MLP framework. MLPs can be employed at the full image resolution to capture fine-grained long-range dependence, which has been proven to be crucial for medical dense prediction tasks and produced consistent improvements in medical image restoration, registration, and segmentation. Moreover, our ablation studies suggest that the choice of different MLP blocks is not of vital importance while employing MLPs at the full image resolution is the key for improving medical dense prediction. We hope this study can encourage the research community to realize the superiority of MLPs over CNNs and transformers for medical dense prediction.

# Supplementary Materials

## A. Dataset and Preprocessing Details

**Restoration.** In the dataset provided by the Ultra 2022 challenge [54], all the data was acquired in a list mode allowing for the rebinding of data to simulate different acquisition times. Therefore, the low-dose PET images can be simulated with a certain dose reduction factor (DRF), which was reconstructed from the counts of a time window resampled at the middle of the acquisition with reduced time. Low-dose PET images were provided with DRF at 4, 10, 20, 50, and 100. All these low-dose PET images are aligned with the corresponding standard-dose PET image. The standard-dose and simulated low-dose PET images are exemplified in Figure S1. The PET images acquired from the SBVQ scanners have a size of 440×440×644 with a voxel spacing of 1.65×1.65×1.65mm$^3$, while the PET images acquired from the UIU scanners have a size of 360×360×673 with a voxel spacing of 1.667×1.667×2.886 mm$^3$. All PET images were converted into standardized uptake value (SUV) maps and then were cropped into overlapping patches with 256×256×32 voxels as the model input. The final restored images were obtained by merging the results of all overlapping patches.

**Registration.** The training set consists of 2,656 brain MRI images acquired from four unlabeled public datasets, including 121 images from ADNI [60], 1087 images from ABIDE [61], 970 images from ADHD [62], and 478 images from IXI [63]. The testing sets are acquired from the Mindboggle [64] and Buckner [65] datasets, where 62 and 110 brain anatomical structures were segmented as labels (exemplified in Figure S2). We conducted standard brain MRI preprocessing procedures, including brain extraction, intensity normalization, and affine registration, with FreeSurfer [65] and FLIRT [73]. All images were affine-transformed and resampled to align with the MNI-152 brain template [74] with 1mm$^3$ isotropic voxels, which were then cropped into 144×192×160 voxels.

**Segmentation.** For head and neck tumor segmentation in the HECKTOR dataset [66], we resampled PET-CT images into isotropic voxels where 1 voxel corresponds to 1mm$^3$. Following [75, 76], each image was cropped to 160×160×160 voxels with the tumor located in the center. PET images were standardized by Z-score normalization, while CT images were clipped to [−1024, 1024] and then mapped to [−1, 1]. For cardiac image segmentation, the ACDC dataset [67] provides cine-MRI images containing tens of 3D MRI frames acquired from different time-points. Only the End-Diastole (ED) and End-Systole (ES) frames of each cine-MRI image have segmentation labels and were adopted in this study. We resampled ED and ES frames into a voxel spacing of 1.5×1.5×3.15mm$^3$ and cropped them to 128×128×32 voxels around the center. The voxel intensity was normalized to range [0, 1] by max-min normalization. The images in the HECKTOR and ACDC datasets are exemplified in Figure S3.

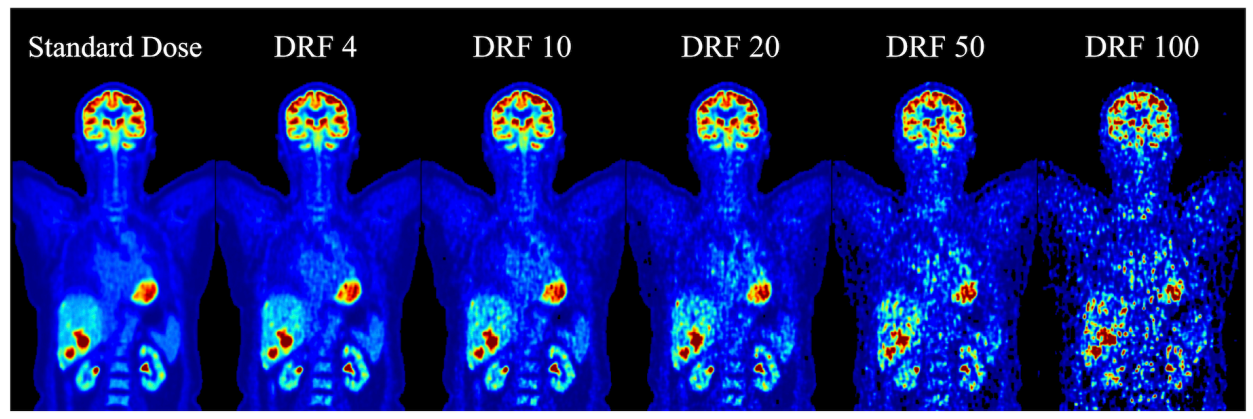

Figure S1: Examples of a standard-dose PET image and low-dose PET images with different dose reduction factors (DRF). The single-channel PET images are colored for illustration, where red and blue indicate high and low image intensity.

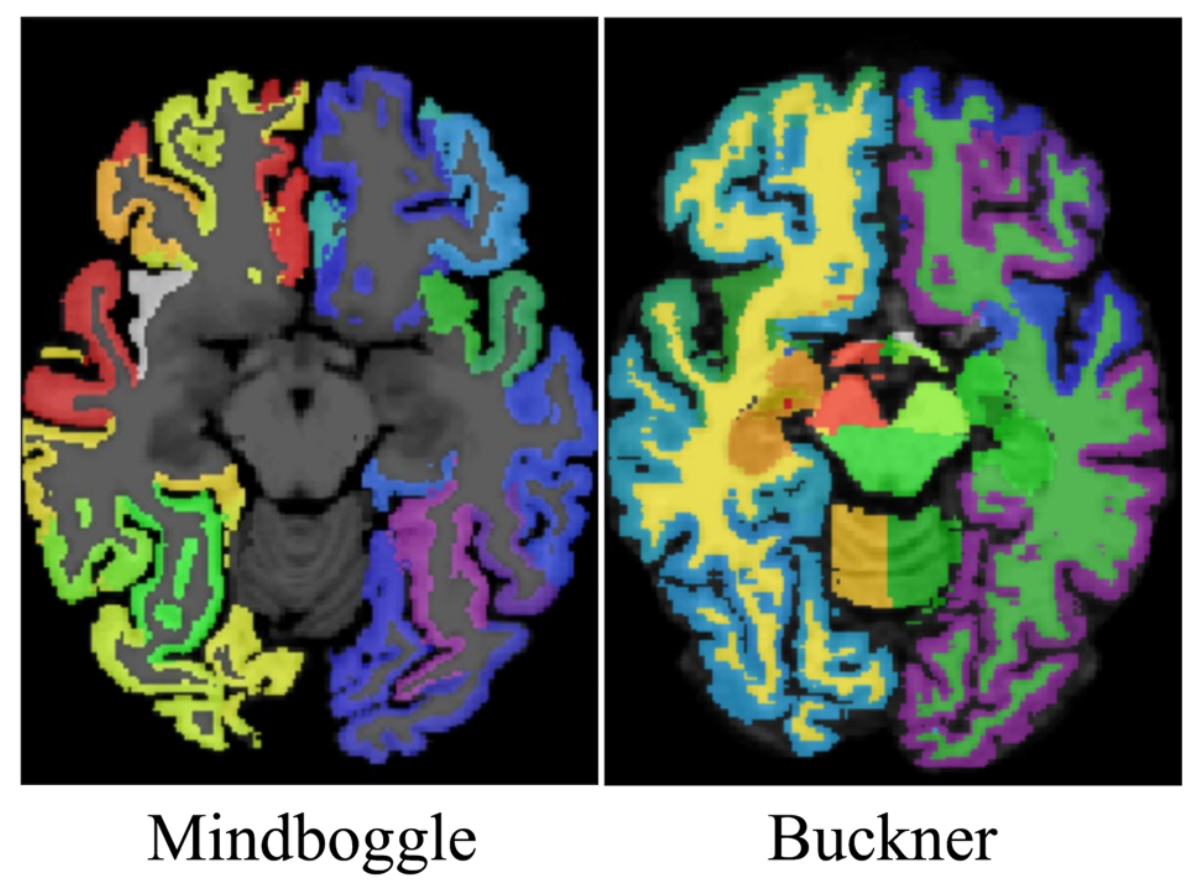

Figure S2: Examples of brain MRI images in the Mindboggle and Buckner datasets, where 62 and 110 labeled brain anatomical structures are colored for illustration.

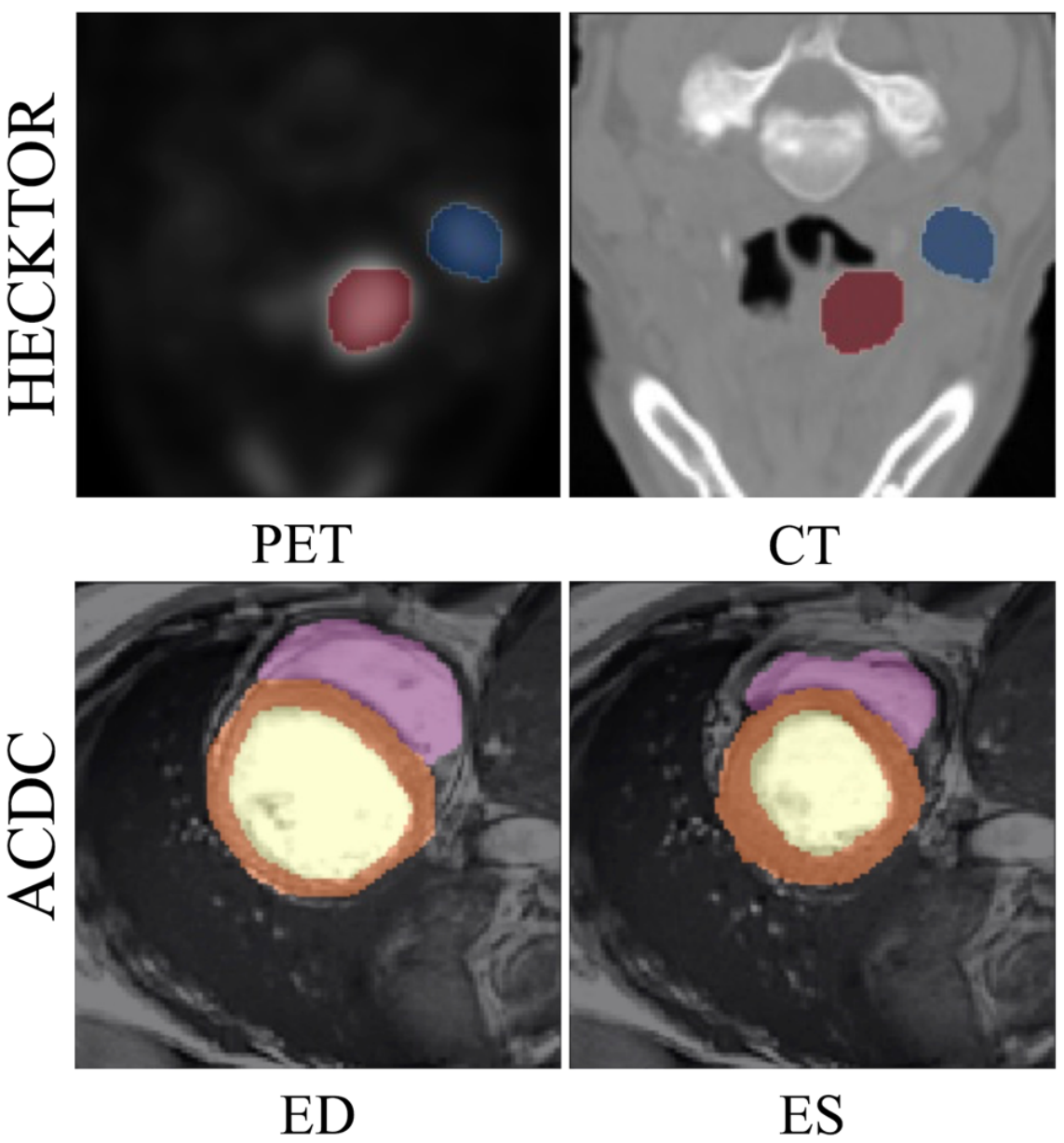

Figure S3: Examples of the images in the HECKTOR (upper) and ACDC (bottom) datasets. The labeled tumors or anatomical regions are colored for illustration.

## B. Definitions of Evaluation Metrics

**Restoration.** NRMSE measures the normalized average difference between the predicted and ground-truth images, which is defined as follows:

$$NRMSE(x,y)=\sqrt{\frac{1}{N}\sum_{i}^{N}\left(\frac{x_i-y_i}{\max(y)-\min(y)}\right)^2}, \quad (1)$$

where the $N$ is the image size, the $x$ is the predicted images, and the $y$ is the ground-truth image. Moreover, SSIM measures the similarity between two images. The formula involves three comparison terms: luminance, contrast, and structure. The SSIM index is calculated as the product of these three terms:

$$SSIM(x,y)=\frac{(2\mu_x\mu_y+C_1)(2\sigma_{xy}+C_2)}{(\mu_x^2+\mu_y^2+C_1)(\sigma_x^2+\sigma_y^2+C_2)}, \quad (2)$$

where the $\mu_x$ and $\mu_y$ are the means of the compared images $x$ and $y$, $\sigma_x^2$ and $\sigma_y^2$ are the variances, $\sigma_{xy}$ is the covariance, $C_1$ and $C_2$ are constants to stabilize the division. Further, PSNR measures the ratio between the maximum possible power of a signal and the power of corrupting noise that affects the quality of its representation. It is often used to assess image quality by comparing it to a reference image. The equation for PSNR in decibels (dB) is:

$$PSNR(x,y)=10\cdot\log_{10}\left(\frac{max(y)}{\frac{1}{N}\sum_{i}^{N}(x_i-y_i)^2}\right), \quad (3)$$

where the $N$ is the image size, the $x$ is the predicted images, and the $y$ is the ground-truth image. Finally, following [54], an average evaluation score $Score_{avg}$ was used to evaluate the overall performance on all low-dose PET images, where different weights were applied to the scores of low-dose PET images with different DRFs:

$$\begin{aligned}Score_{avg} &= w_1 * score_{DRF100} + w_2 * score_{DRF50} \\ &+ w_3 * score_{DRF20} + w_4 * score_{DRF10} \\ &+ w_5 * score_{DRF4},\end{aligned} \quad (4)$$

where the *score* can be either of the evaluation metrics, and the $w_1$, $w_2$, $w_3$, $w_4$ and $w_5$ are the average weights being set as 0.35, 0.25, 0.2, 0.15, and 0.5, respectively.

**Registration.** The registration accuracy was evaluated using the Dice similarity coefficients (DSC) between the segmentation labels of the fixed images $I_f$ and the warped image $I_{m\circ\phi}$, which is defined as follows:

$$DSC(S_f,S_{m\circ\phi})=\frac{1}{N}\sum_{i}^{N}\frac{2\left|S_f^i\cap S_{m\circ\phi}^i\right|}{\left|S_f^i\right|+\left|S_{m\circ\phi}^i\right|}, \quad (5)$$

where the $N$ is the number of the labeled anatomical structures, the $S_f$ and $S_{m\circ\phi}$ is the segmentation labels of $I_f$ and $I_{m\circ\phi}$, and $I_{m\circ\phi}$ and $S_{m\circ\phi}$ is derived by warping the $I_m$ and $S_m$ with the produced spatial transformation $\phi$. The smoothness and invertibility of the spatial transformation $\phi$ were evaluated via the percentage of negative Jacobian determinants (NJD), which is defined as follows:

$$NJD(\phi)=\frac{1}{H\times W\times D}\sum_{p}^{H\times W\times D}(Jdet(\phi_p)\leq 0), \quad (6)$$

where $(H, W, D)$ is the shape of images and the $Jdet(\phi_p)$ is the Jacobian determinant of $\phi$ in pixel location $p$.

**Segmentation.** DSC and HD95 were used to evaluate the consistency between the segmentation prediction $S_P$ and the ground truth label $S_L$, which are defined below:

$$DSC(S_P,S_L)=\frac{1}{N}\sum_{i}^{N}\frac{2\left|S_P^i\cap S_L^i\right|}{\left|S_P^i\right|+\left|S_L^i\right|}, \quad (7)$$

$$\begin{aligned}&HD95(S_P,S_L)=\\ &P_{95}\left\{sup_{i\in S_P}inf_{i\in S_L}d(i,j), sup_{i\in S_L}inf_{i\in S_L}d(i,j)\right\},\end{aligned} \quad (8)$$

where the $N$ is the number of the labeled region, the $d(i,j)$ is the Euclidean distance between the points $i$ and $j$, sup and inf are the supremum and infimum respectively, and the $P_{95}$ is the 95[th] percentile.

## C. Loss Functions

**Restoration.** We adopt the adversarial strategy of GAN and modify it to incorporate both voxel-wise content loss and image-wise adversarial loss, so as to maximize the alignment between the restored images and ground-truth standard-dose images. The generator $P$ is a restoration network (e.g., MLPUnet), while the discriminator $D$ takes the fake (restored) or real image pairs of low-dose and standard-dose PET images as input and then differentiates whether they are real or fake. The discriminator $D$ contains five convolution blocks, and each block is composed of a 2-strided convolutional layer, LeakyReLU activation, and batch normalization. The activation of the last convolution block is replaced by a sigmoid function to produce a probability map that should be close to 0 (for fake image pairs) or 1 (for real image pairs). The voxel-wise content loss and adversarial loss are formulated as follows:

$$\mathcal{L}_{content} = \mathrm{E}_{V_L\sim P_L}[\|V_S-P(V_L)\|_1], \quad (9)$$

$$\begin{aligned}\mathcal{L}_{adv} &= E_{V_L\sim P_L,V_S\sim P_S}[(D(V_L,V_S)-1)^2] \\ &+E_{V_L\sim P_L}[D(V_L,P(V_L))^2],\end{aligned} \quad (10)$$

where the $V_L$ and $V_s$ are the low-dose and standard-dose PET images. The final restoration loss $\mathcal{L}_{Res}$ is composed of the content loss $\mathcal{L}_{content}$ and the adversarial loss $\mathcal{L}_{adv}$, which is defined as follows:

$$\mathcal{L}_{Res} = \lambda_1\mathcal{L}_{content}+\lambda_2\mathcal{L}_{adv}, \quad (11)$$

where we empirically set $\lambda_1$=300 and $\lambda_2$=1.

**Registration.** The registration networks are optimized using an unsupervised loss $\mathcal{L}_{Reg}$ without requiring labels. The $\mathcal{L}_{Reg}$ consists of two terms $\mathcal{L}_s$ and $\mathcal{L}_r$, where the $\mathcal{L}_s$ is an image similarity term that penalizes the differences between the warped image $I_{m\circ\psi}$ and the fixed image $I_f$, while the $\mathcal{L}_r$ is a regularization term encouraging smooth and realistic spatial transformations $\phi$. For the $\mathcal{L}_s$, we adopted the negative local normalized cross-correlation (NCC). Specifically, let $\hat{I}(p)$ denote the local mean intensity of image $I$ in the location $p$:

$$\hat{I}(p)=\frac{1}{n^3}\sum_{p_i}I(p_i), \quad (12)$$

where the $p_i$ iterates over a $n^3$ neighboring region around $p$, with $n=9$ in our experiments. Then, the $\mathcal{L}_s$ between the $I_f$ and $I_{m\circ\psi}$ is defined as:

$$\mathcal{L}_s\left(I_f, I_{m\circ\psi}\right) = -\sum_{p\in\Omega} \frac{\left(\sum_{p_i}[I_f(p_i)-\hat{I}_f(p)][I_{m\circ\psi}(p_i)-\hat{I}_{m\circ\psi}(p)]\right)^2}{\left(\sum_{p_i}[I_f(p_i)-\hat{I}_f(p)]^2\right)\left(\sum_{p_i}[I_{m\circ\psi}(p_i)-\hat{I}_{m\circ\psi}(p)]^2\right)}, \quad (13)$$

where the $\Omega$ denotes the whole image space. For the $\mathcal{L}_r$, a diffusion regularizer was imposed on the $\phi$ to encourage its smoothness:

$$\mathcal{L}_r(\phi) = \sum_{p\in\Omega} ||\nabla\phi(p)||^2, \quad (14)$$

where the $\nabla$ is the spatial gradient operator. Finally, the registration loss $\mathcal{L}_{Reg}$ is defined as:

$$\mathcal{L}_{Reg} = \mathcal{L}_s\left(I_f, I_{m\circ\psi}\right) + \lambda\mathcal{L}_r(\phi), \quad (15)$$

where the $\lambda$ is a regularization parameter balancing the registration accuracy and transformation smoothness, which is set as 1 in our experiments.

**Segmentation.** The segmentation loss $\mathcal{L}_{seg}$ is the sum of the Dice loss $\mathcal{L}_{Dice}$ and the Focal loss $\mathcal{L}_{Focal}$, which is defined as follows:

$$\mathcal{L}_{Seg} = \mathcal{L}_{Dice} + \mathcal{L}_{Focal}, \quad (16)$$

$$\mathcal{L}_{Dice} = \frac{1}{N}\sum_i^N \frac{2\sum_{j\in\Omega} P_{ij}G_{ij}}{\sum_{j\in\Omega} {P_{ij}}^2 + \sum_{j\in\Omega} {G_{ij}}^2}, \quad (17)$$

$$\mathcal{L}_{Focal} = -\frac{1}{N}\sum_i^N \sum_{j\in\Omega}[\alpha G_{ij}\left(1-P_{ij}\right)^\gamma \log\left(P_{ij}\right) - (1-G_{ij}){P_{ij}}^\gamma \log(1-P_{ij})], \quad (18)$$

where the $N$ is the number of the labeled region, the $\Omega$ denotes the whole image space, the $P$ is the output of the segmentation network, and the $G$ is the ground-truth labels. In the $\mathcal{L}_{Focal}$, the $\alpha$ is a parameter for the trade-off between precision and recall while the $\gamma$ is a focusing parameter, which are set to 0.25 and 2 in our experiments.

## D. Implementation Details

**Restoration.** The restoration networks were trained for 100 epochs using an Adam optimization with a batch size of 4. The learning rate was initially set as 2e-4, which was then linearly decreased with a factor of 0.1 and patience of 5 epochs. An early stopping strategy was used to avoid overfitting, which terminated the training process when the learning rate exceeded 2e-6.

**Registration.** We adopted an ADAM optimizer with a learning rate of 0.0001 and a batch size of 2. The networks were trained for a total of 100,000 iterations with inter-patient image pairs randomly picked from the training set. Validation was performed after every 1,000 iterations and the weights achieving the highest validation result were preserved for final testing.

**Segmentation.** The segmentation networks were trained for 12,000 iterations using an Adam optimizer with a batch size of 4. The learning rate was set as 1e−4 initially and then reset to 5e−5 and 1e−5 at the 4,000th and 8,000th training iterations. Data augmentation was applied in real-time during training to minimize overfitting, including random affine transformations and additional random cropping to 128×128×128 voxels only for the HECKTOR dataset. Validation was performed after every 200 training iterations and the weights achieving the highest validation result were preserved for final testing.

## E. Settings of Ablation Studies

In the first ablation study, the Swin transformer blocks and the residual convolution blocks have the same feature dimensions/channels as the original MLP blocks at each stage. For Swin transformer blocks, the two MLP blocks at each stage were replaced by two Swin transformer blocks with window- and shifted window-based multi-head self-attention (W-MSA and SW-MSA). The Swin transformer blocks have a window size of 7×7×7 and a head number of feature dimensions/16. For residual convolution blocks, the two MLP blocks at each stage were replaced by two convolutional layers with a kernel size of 3×3×3 and a shortcut connection between the input and output, followed by batch normalization and ReLU activation. Note that, for a fair comparison, channel attention blocks were equally employed following the MLP, transformer, or convolution blocks, which consist of layer normalization, convolutional layers, LeakyReLU activation, and squeeze-and-excitation (SE) channel attention [77], with a shortcut connection.

In the second ablation study, various MLP blocks were employed in the full-resolution MLP framework. Similarly, for a fair comparison, these different MLP blocks have the same feature dimensions as the original settings in the framework, and channel attention blocks were also used following each MLP block. The hyperparameter settings of all MLP blocks are detailed in Supplementary Section F.

## F. Settings of MLP Blocks

The hyperparameter settings of MLP blocks are listed in Table S1. The Multi-axis gated MLP blocks were used as default in all experiments, while other MLP blocks were used merely in the second ablation study.

| MLP block | Hyperparameter settings |
|---|---|
| Hire-MLP | Inner-region size = [8, 8, 6, 6, 4]<br>Cross-region size = [4, 4, 3, 3, 2] |
| sMLP | Hyperparameter-free<br>Global mixing along spatial axes |
| Swin-MLP | Window size = [8, 8, 6, 6, 4]<br>Shift size = [4, 4, 3, 3, 2] |
| Multi-axis gated MLP | Block size = [8, 8, 6, 6, 4]<br>Grid size = [8, 8, 6, 6, 4] |

Table S1: Hyperparameter settings of the MLP block used in the experiments. The list $[n_1, n_2, n_3, n_4, n_5]$ provides the settings in the five stages of the full-resolution framework.

## G. Architecture of Task-specific Decoders

For restoration and segmentation, the decoder of Swin-UNETR [12] is adopted, which is illustrated in Figure S4. The multi-scale features derived from the full-resolution

MLP framework, $[F_1, F_2, F_3, F_4, F_5]$, are first processed by a residual convolution block independently, and then the resultant features are fed into a progressively upsampling decoder with residual convolution blocks. Each residual convolution block consists of two successive convolutional layers with a kernel size of 3×3×3, followed by batch normalization and ReLU activation. A shortcut connection or 1×1×1 convolution is employed between the input and output of each residual convolution block.

For registration, the decoder of TransMorph [10] is adopted, which is illustrated in Figure S5. The multi-scale features $[F_1, F_2, F_3, F_4, F_5]$ are directly fed into a progressively upsampling decoder with simple convolution blocks. Each convolution block consists of two successive convolutional layers with a kernel size of 3×3×3, followed by batch normalization and LeakyReLU activation.

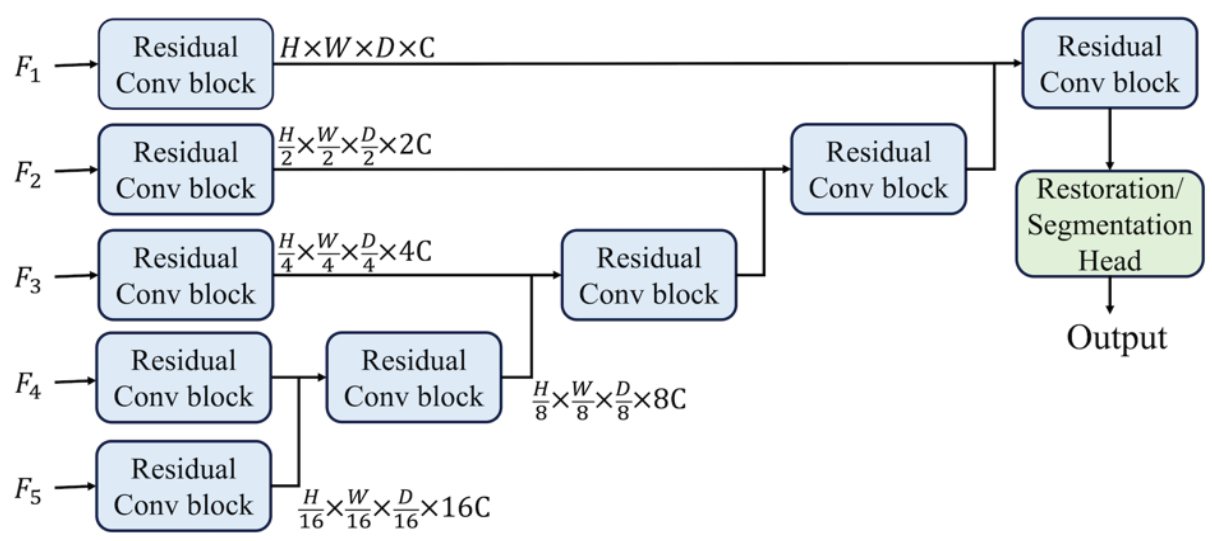

Figure S4: The decoder for restoration and segmentation.

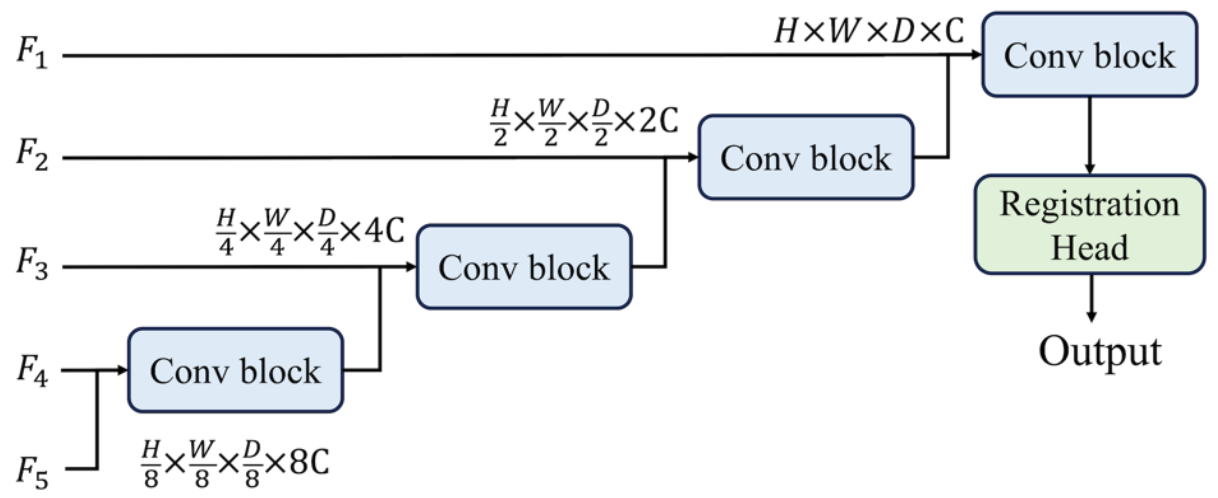

Figure S5: The architecture of registration decoder.